\documentclass[a4paper, prb, twocolumn, showpacs]{revtex4}

\usepackage{amssymb}
\usepackage{amsmath}
\usepackage{color}
\usepackage{graphics, graphicx}
\usepackage{bbold}
\usepackage{psfrag}
\usepackage{mathcomp}

\newcommand{\Ham} {\hat{H}}

\renewcommand{\ao} {\hat{b}}

\newcommand{\oao} {\hat{a}}
\newcommand{\oco} {\hat{a}^\dagger}

\newcommand{\io} {i\omega_n}

\begin{document}

\author{A. Hubener$^{1}$}
\author{M. Snoek$^{1,2}$}
\author{W. Hofstetter$^{1}$}
\affiliation{$^{1}$Institut f\"ur Theoretische Physik, Johann Wolfgang
Goethe-Universit\"at, 60438 Frankfurt am Main, Germany \\ $^{2}$ Institute for Theoretical
Physics, Valckenierstraat 65, 1018 XE Amsterdam, the Netherlands}


\title{Magnetic Phases of Two-Component Ultracold Bosons in an Optical Lattice}

\begin{abstract}
We investigate spin-order of ultracold bosons in an optical lattice by means of Dynamical
Mean-Field Theory.   
A rich phase diagram with anisotropic magnetic order is found, both for the ground state
and at finite temperatures. 
Within the Mott insulator, a ferromagnetic to antiferromagnetic transition can be tuned
using a spin-dependent 
optical lattice. In addition we find a supersolid phase, in which superfluidity coexists
with antiferromagnetic spin order. 
We present detailed phase diagrams at finite temperature for the experimentally realized 
heteronuclear $^{87}$Rb - $^{41}$K mixture in a three-dimensional optical lattice.
\end{abstract}

\maketitle

\section{Introduction} 
Ultracold atoms in optical lattices give access to studies of quantum magnetism with
unprecedented control and precision. 
Whereas efficient cooling of fermionic atoms in optical lattices remains an experimental
challenge, 
mixtures of bosonic atoms can more easily be cooled to the relevant temperature scales, 
thus realizing the Bose-Hubbard Model \cite{Fisher, Jaksch} with multiple species. Indeed,
recent experiments have already succeeded in loading a heteronuclear mixture into an
optical lattice \cite{Minardi08}. The interaction between the two species has been
adressed by means of a Feshbach resonance \cite{Minardi08b}, which offers a direct way of
mapping out the phase diagram as a function of the interspecies interaction strength. 
Moreover, superexchange processes have been directly observed in a mixture of
two-component bosons \cite{Lukin08}. 

Whereas the phase diagram of spinless bosons is qualitatively well captured by Gutzwiller
Mean-Field theory \cite{Gutzwiller}, for a multispecies system dynamical correlations are
important, because they give rise to spin order. 
Previous theoretical studies have either discussed the weak tunneling limit
\cite{Kuklov03,Duan03} or 
performed expansions around mean-field \cite{Altman03} or the strong-coupling limit
\cite{Isacsson05, Powell09}. Numerical non-perturbative methods are available in one spatial dimension \cite{Alon06, Kleine08}, and   
Quantum Monte-Carlo simulations were very recently performed for two spatial dimensions
\cite{QMC2d}. 
In this article we focus on the generic three-dimensional situation where Dynamical
Mean-Field Theory (DMFT) has been established \cite{DMFT} as a highly reliable,
nonperturbative approach to strongly correlated fermionic quantum systems. 
Here we apply the bosonic version of DMFT (BDMFT), as recently introduced by Byczuk and Vollhardt
\cite{Byczuk08}. 
BDMFT treats condensed and normal bosons on equal footing.
It is non-perturbative and hence can be applied within the full range from small to large couplings. The control parameter is the lattice coordination number $z$, i.e. the theory becomes exact in infinite dimensions. 
We present the BDMFT equations as a controlled $1/z$ expansion, up to subleading order. 
Our derivation is thus different from the original proposal \cite{Byczuk08}, in which BMDFT is 
constructed as a well-defined theory in strictly infinite dimensions, which requires  
a different scaling of superfluid and normal parts of the action. 
In contrast, our derivation is based on a uniform scaling $\sim 1/z$ of the bosonic hopping amplitude. 
To leading order this yields Gutzwiller Mean-Field theory \cite{Gutzwiller}, while from the subleading terms 
of order $O(1/z)$ we obtain the BDMFT equations. We thus regard BDMFT as an expansion in $1/z$ 
around Gutzwiller, which in our opinion is the most natural viewpoint. 

In practice, BDMFT turns out to be an efficient and fast scheme, which allows to map out phase diagrams with high resolution. Moreover, it not only allows for non-perturbative calculation of local observables, but also the spectral function, relevant for RF-spectroscopy, is directly accessible.

While in Ref.~\onlinecite{Byczuk08} calculations were only performed for a simplified lattice model
with 
partially immobile bosons, here we apply bosonic DMFT to the full two-component Bose-Hubbard model in finite spatial dimensions. 
Besides the
superfluid, we identify $XY$-ferromagnetic and $Z$-antiferromagnetic phases, in which 
translational symmetry is spontaneously broken and anisotropic magnetic order arises. 
Moreover, we find a supersolid phase where superfluidity coexists with antiferromagnetic
spin order. We investigate the stability of these phases against thermal fluctuations,
paying special attention to the experimentally relevant case of a heteronuclear $^{87}$Rb
- $^{41}$K mixture in a three-dimensional optical lattice \cite{Minardi08}. 

\section{Method} 
We first describe how BDMFT can be
implemented for the Bose-Hubbard Hamiltonian. In particular we identify an Anderson
Hamiltonian that  reproduces the local effective action and allows us to solve the BDMFT self-consistency problem. 

The starting point for our investigation is the 
 multispecies single-band Hubbard model within tight-binding
approximation
\begin{equation}
\hat{H} = -\hspace{-2pt}\sum_{\langle ij\rangle , \nu }\hspace{-2pt} \left(t_\nu
\hat{b}^{\dagger}_{i\nu}\hat{b}^{}_{j\nu} + \text{h.c.} \right)
	+\frac{1}{2} \sum_{i,\mu\nu} U_{\nu\mu} \hat{n}^{}_{i\nu} \,\Big(
\hat{n}^{}_{i\mu} - \delta_{\nu\mu} \Big),
\end{equation}
 which provides an accurate description of bosonic atoms in a sufficiently strong optical lattice. Here
 $t_\nu$ are (species-dependent) hopping amplitudes,  $U_{\nu\mu}$ contains the inter- and
 intraspecies interactions and $\langle i j \rangle$ indicates a summation over nearest
 neighbor sites $i$ and $j$.
 In the spirit of the cavity derivation of the fermionic DMFT equations \cite{DMFT}
 we consider a single lattice site (called the ``impurity site'') and formally integrate
 out all the other degrees of freedom. This defines the effective action of the impurity
 site as
\begin{equation}
	Z_\text{imp} = \frac{Z}{Z^{(0)}} = \int \prod_{\nu} {\cal D} b^{*}_{0,\nu} \,
{\cal D} b^{}_{0,\nu} \,e^{-S_\text{imp}}, 
\end{equation}
where $Z$ is the full partition function and $Z^{(0)}$ is the partition function of the
cavity system without the impurity.
For reasons of brevity we derive the effective impurity action and perform the numerical calculations in this article for the case of a Bethe
lattice (Cayley tree), which in infinite dimensions ($z \rightarrow \infty$) has a
semicircular density of states \cite{DMFT}: 
\mbox{$\rho_0 (\epsilon) =  \sqrt{4zt^2-\epsilon^2}/2 \pi z t^2$}. 
The use of the semicircular DOS in our calculations has merely technical reasons because
this choice simplifies the DMFT equations. Our obtained results remain qualitatively
similar for any symmetric DOS representing a bipartite lattice. This is in particular true for the three-dimensional cubic lattice, which has only mild Van Hove singularities. For fermionic DMFT it has been established that the agreement between results on the Bethe lattice and the cubic lattice is not only qualitative, but also quantitative, with a typical accuracy of around ten percent. We find that the same is true for BDMFT, as we show below for the case of single component bosons, where we compare the BDMFT results with numerically exact Quantum Monte Carlo results.

In deriving the effective impurity action, we first formally rescale all hopping parameters as $t_\nu = t_\nu^*/z$, such that $1/z$
appears as the small parameter in the theory.
 Based on the linked cluster theorem, the action of the impurity site up to subleading
order  in $1/z$ is then obtained in the standard way \cite{DMFT} as:
\begin{widetext}
\begin{eqnarray}
S_\text{imp} \hspace{-0.2cm}
	&= & \hspace{-0.25cm} \int_0^\beta \hspace{-0.2cm} d \tau d\tau' \sum_{\mu\nu}
 \Bigg( \hspace{-0.1cm} \begin{array}{c} b^*_{0\mu} (\tau) \\ b_{0\mu} (\tau) \end{array}
\hspace{-0.1cm} \Bigg)^{\hspace{-0.1cm}T}
			\hspace{-0.05cm} \left(\begin{array}{cc} \hspace{-0.1cm}
				(\partial_{\tau'}-\mu_\mu)\delta_{\mu \nu}+ t_\mu t_\nu
\hspace{-0.25cm} \sum \limits_{\langle 0i\rangle,\langle 0j\rangle} \hspace{-0.25cm}
G_{\mu \nu, ij}^1 (\tau, \tau') 
			&  t_\mu t_\nu  \hspace{-0.25cm} \sum \limits_{\langle
0i\rangle,\langle 0j\rangle} \hspace{-0.25cm}   G^2_{\mu\nu, ij}(\tau, \tau') \\
			 t_\mu t_\nu  \hspace{-0.25cm} \sum \limits_{\langle
0i\rangle,\langle 0j\rangle} \hspace{-0.25cm} {G^2_{\mu\nu, ij}}^*(\tau', \tau)
	 & (-\partial_{\tau'}-\mu_\mu)\delta_{\mu \nu}+ t_\mu t_\nu   \hspace{-0.25cm}
\sum \limits_{\langle 0i\rangle,\langle 0j\rangle} \hspace{-0.25cm} G_{\mu \nu, ij}^1
(\tau', \tau)
			 \hspace{-0.1cm} \end{array}\right) \hspace{-0.15cm}		
				\Bigg(\begin{array}{c} \hspace{-0.1cm} b_{0\nu} (\tau') \\
b^*_{0\nu} (\tau') \end{array} \hspace{-0.1cm}\Bigg)	\nonumber\\
 && + \quad\int_0^\beta  d\tau \left\{ \frac{1}{2}\sum_{\nu\mu} U_{\nu\mu}\;
n_{0\nu}(\tau) \,\Big( n_{0\mu}(\tau) - \delta_{\nu\mu} \Big)
 	 - \sum_{\langle 0i\rangle,\nu } t_\nu \Big( b^{*}_{0\nu}(\tau)
\phi^{}_{i,\nu}(\tau) 
 		+ b_{0\nu}(\tau) \phi^{*}_{i,\nu}(\tau) \Big) \right\}
\label{effaction}
\end{eqnarray}
\end{widetext}
Here we have defined  
\begin{equation}
\phi^{}_{i,\nu}(\tau) = \langle b_{i, \nu} (\tau)
\rangle_0 
\end{equation}
as the superfluid order parameters, and  
\begin{eqnarray}
\hspace{-0.5cm}G_{\mu \nu, ij}^1 (\tau, \tau')\hspace{-0.2cm}&=&\hspace{-0.2cm}- \langle b_{i, \mu} (\tau) b_{j, \nu}^* (\tau')
\rangle_0 + \phi_{i, \nu} (\tau) \phi_{j, \mu}^* (\tau'), \\
\hspace{-0.5cm}G_{\mu \nu, ij}^2 (\tau, \tau')\hspace{-0.2cm}&=&\hspace{-0.2cm}- \langle b_{i, \mu} (\tau) b_{j, \nu} (\tau')
\rangle_0 + \phi_{i, \nu} (\tau) \phi_{j, \mu} (\tau')
\end{eqnarray} 
as the diagonal and off-diagonal parts of the connected Green's
functions, respectively. The notation $\langle \ldots \rangle_0$ means that the expectation value is
taken in the cavity system excluding the impurity site. For finite $z$ the action
($\ref{effaction}$) coincides with the one previously derived in \cite{Byczuk08}. Note
however, that our derivation is different. In the original proposal \cite{Byczuk08}, 
bosonic Dynamical Mean-Field Theory (BMDFT) is 
constructed as a well-defined theory in strictly infinite dimensions, which requires  
different scaling of superfluid and normal parts of the action. 
In contrast, our derivation is based on a uniform scaling $\sim 1/z$ of the bosonic
hopping amplitude, since we focus on finite dimensions and and our goal is to make direct contact with the three-dimensional experimental situation. 
The terms involving Green's functions in the action (\ref{effaction}) are of order $O(1/z)$, since they come with two factors of $t_\nu \sim 1/z$ and one summation over neighboring sites, which gives a factor $z$. All the other terms are of order $O(1)$; for the last term in the action this follows from the fact that it involves one factor of $t_\nu \sim 1/z$ which cancels against the factor $z$ arising from the summation over neighboring sites. 
Therefore to leading order the action (\ref{effaction}) yields Gutzwiller Mean-Field theory \cite{Gutzwiller}, while by including 
the subleading terms 
of order $O(1/z)$ we obtain the BDMFT equations. Hence we regard BDMFT as an expansion in
$1/z$ around Gutzwiller; this is in our opinion the most natural viewpoint.

\begin{figure}
\includegraphics[scale=.3]{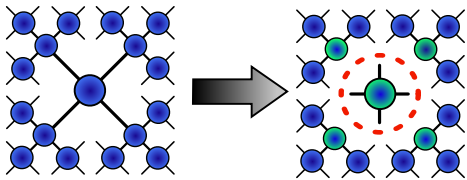}
\caption{Illustration of the cavity method. Sites which are connected to the impurity
(coloured greenish)  have one neighbor less once the impurity is removed.}
\label{bethe}
\end{figure}

To proceed, expectation values in the cavity system need to be identified with those
on the impurity site, in order to obtain a closed self-consistency loop.  
Since sites at the edge of the
cavity have one neighbor less compared to the impurity site (see Fig.~\ref{bethe}), simply identifying the expectation values yields an error of
order $1/z$. 
For the Green's functions this poses no
problem, because they already appear at subleading order in the action, but it yields a relevant correction to the superfluid order parameter and turns out to be essential for quantitatively accurate predictions of the phase diagram. Details regarding the implementation will be given below.

We now turn to the solution of the effective action. This we do in the spirit of the exact
diagonalization (ED) solution of fermionic DMFT \cite{DMFT}.
We represent the effective action (\ref{effaction}) by an Anderson Impurity Hamiltonian
$\Ham_A$:

\begin{eqnarray}
\hat{H}_A &=& - \sum_\nu zt_\nu \big(\phi_\nu^*\ao_\nu + {\rm h.c.} \big) + \frac{1}{2}
\sum_{\mu\nu} U_{\nu\mu} \hat{n}^{}_{\nu} \big( \hat{n}^{}_{\mu} - \delta_{\nu\mu} \big)
\nonumber \\ && - \sum_\nu \mu_\nu \hat{n}_\nu \nonumber
	 +\sum_{l}  \epsilon_l \oco_l\oao_l \\&& 
	+ \sum_{l,\nu} \Big( V_{\nu,l} \oco_l\ao_\nu + W_{\nu,l} \oao_l\ao_\nu + {\rm
h.c.} \Big) 
\end{eqnarray}

The chemical potential and interaction term are directly inherited from the Hubbard
Hamiltonian. The Gutzwiller term represents the bath of condensed bosons with
superfluid order parameters $\phi_\nu$ for every component. The bath of normal bosons is
modeled by a finite number of orbitals with creation operators $\oao_l^\dagger$ and energies
$\epsilon_l$. These orbitals are coupled to the impurity via normal-hopping amplitudes
$V_{\nu, l}$ and anomalous-hopping amplitudes $W_{\nu, l}$. The anomalous hopping terms are
needed to generate the off-diagonal elements of the hybridization function.
We define the
following hybridization functions:
\begin{eqnarray}
\Delta_{\nu\mu}^1(i \omega_n) & \equiv & -\sum_l
\frac{V_{\nu,l}V_{\mu,l}}{\epsilon_l-\io} + \frac{W_{\nu,l}W_{\mu,l}}{\epsilon_l+\io} \\
\Delta_{\nu\mu}^2(i \omega_n) & \equiv &  -\sum_l
\frac{V_{\nu,l}W_{\mu,l}}{\epsilon_l-\io} + 
\frac{V_{\nu,l}W_{\mu,l}}{\epsilon_l+\io}
\end{eqnarray}
Integrating out the orbitals leads to the same effective action (3), if the following
identification is made:
\begin{eqnarray}
zt_\nu t_\mu G^{1,2}_{\nu\mu} (i \omega_n) & \hat{=} & 
\Delta_{\nu\mu}^{1,2} (i\omega_n).
\label{ssG}
\end{eqnarray}
These self-consistency conditions are completed by the condition for the superfluid order parameter 
\begin{equation}
\phi_\nu =
\langle\ao_\nu\rangle_{0}^{z-1}.
\label{ssphi}
\end{equation}
The notation $\langle \ldots \rangle^{z-1}_0$ means that the expectation value is corrected for the missing neighbor on the sites adjacent to the impurity. Since this is a correction of order $O(1/z)$ and $1/z$ is small, this correction is implemented by means of perturbation theory in $1/z$. 
Equations (\ref{ssG}) and (\ref{ssphi}) thus consitute the set of BDMFT self-consistency conditions. 

The self-consistency loop is solved as follows: starting from an initial choice for the superfluid order parameter and the Anderson parameters,
 the Anderson Hamiltonian is constructed in the Fock basis and diagonalized to obtain the  eigenstates and eigenenergies. New superfluid order parameters are then obtained from $\phi_\nu = \langle\ao_\nu\rangle_{0}^{z-1}$.
The eigenstates and energies also allow us to calculate the Green's functions. Subsequently, new Anderson parameters are obtained by fitting the hybridization functions to their corresponding Green's functions according to Eq. (\ref{ssG}), which is done 
by a conjugate gradient method.
With this new Anderson Hamiltonian the procedure is iterated until convergence is reached.

We note here that this derivation is independent of temperature. This implies that we
cannot only determine ground state properties, but also obtain information about the
thermodynamics of lattice bosons, as we will show in the following results. 
Similar to fermionic DMFT this raises the question how BDMFT deals with situations with broken symmetries, in which case Goldstone modes are present in the spectrum. Indeed, the gapless long wavelength excitations are absent from the DMFT spectrum \cite{DMFT}. However, since in three dimensions the spectral weight of the Goldstone mode is finite and generally small, this approximation can be justified and does not prohibit qualitative agreement between (B)DMFT results and more exact methods (if available), even in symmetry-broken states.

\section{Results}

\subsection{Single component bosons} 
We now first apply BDMFT for the case of single component bosons, in which case we can compare the results with numerically exact Quantum Monte Carlo data \cite{QMC3d} and strong coupling expansions \cite{Pelster, Pelster2} on the cubic lattice, and with the exact solution on the Bethe lattice \cite{Semerjian09}.
\begin{figure}
\includegraphics[scale=.5]{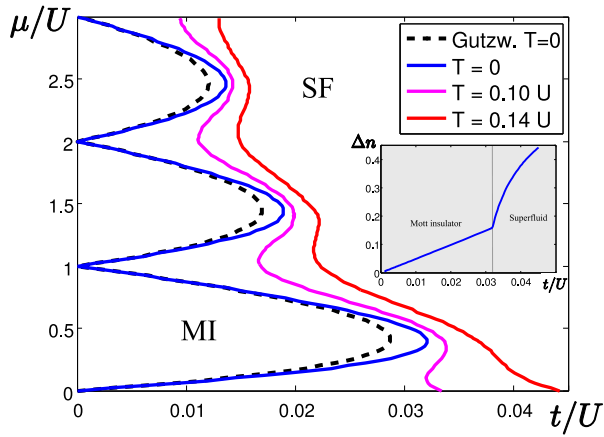}
\caption{Single component phase diagram for various temperatures as obtained by BDMFT. In the inset: number fluctuations for density $n=1$ and $T=0$. The lattice coordination number is chosen as $z=6$.} 
\label{PDspinless}
\end{figure}
Solving the BDMFT equations for the single-component Bose-Hubbard model leads to an extension of
the Mott-insulating lobes compared to the mean-field results (see Fig.~\ref{PDspinless}).
The agreement with the exact results on the Bethe lattice \cite{Semerjian09} is very good: for the lowest Mott lobe the phase boundaries agree within a few percent, whereas for the higher Mott lobes the agreement is even better.
The predicted shift on the cubic lattice is slightly larger \cite{QMC3d, Pelster, Pelster2}, which is due to the different lattice structure. This quantitative agreement with the exact solution clearly shows that the applied $1/z$ expansion is a very good approximation for a three-dimensional system with $z=6$.
It is important to note that to obtain this result the $1/z$ correction of the superfluid order parameter discussed in the previous section is crucial.

Besides this quantitative agreement regarding the boundary of the Mott insulating lobes, it is also important to note that  BDMFT predicts nonzero number fluctuations in the Mott insulator, as shown in the inset of Fig.~\ref{PDspinless}.  The number fluctations are clearly finite in the Mott state and decay as $1/U$ for large $U$. This is in contrast to the Gutzwiller mean-field result, where number fluctuations strictly vanish within the Mott state.
Since these finite fluctuations in the Mott state are essential for resolving spin order in the case of multicomponent mixtures, spin order cannot be resolved withing Gutzwiller. In contrast, the BDMFT contains number fluctuations in the Mott insulating state and is able to describe spin order as we describe in the following subsection.

\subsection{Two-component bosons}
The two-component Bose-Hubbard model has a very rich phase diagram, because additional spin order exists in the Mott phase.
Here we focus on the situation that the total particle density is fixed at one boson per site. In the strong coupling limit, i.e for $t_\nu \ll U_{\mu \nu}$, the system can be mapped to a spin model, which predicts the existence of a $Z$-antiferromagnet
and a $XY$-ferromagnetic phase \cite{Kuklov03, Duan03}. In terms of the particle
creation/annihilation operators $\hat b_1, \hat b_2$, all the insulating phases have the property that $\langle \hat b_1 \rangle = \langle \hat b_2 \rangle = 0$. The $Z$-antiferromagnetic phase breaks the translational symmetry and is
defined by the antiferromagnetic order parameter $\Delta^\mu_{\rm af} = | n_{\alpha, \mu} - n_{\bar
\alpha, \mu}|$ being nonzero, where $\mu$ denotes the component and $\alpha$ ($\bar
\alpha = -\alpha$) the sublattice. The correlator $\langle \hat b_1^\dagger \hat b_2 \rangle$ vanishes in the Z-antiferromagnetic phase. 
The $XY$-ferromagnet is defined by the  local
correlator $\langle \hat b_1^\dagger \hat b_2 \rangle$ 
being nonzero, and is also termed a counter-flow superfluid. This state does not break the translational symmetry and hence $\Delta^\mu_{\rm af} = 0$.

   At weaker coupling the spin model breaks down, because
quantum fluctuations become important. A fluctuation calculation \cite{Altman03} and
strong coupling expansion \cite{Isacsson05} have extended the spin model results to weaker
coupling, where a transition to a superfluid phase $\langle \hat b_1 \rangle, \langle \hat b_2
\rangle \neq 0$ takes place. The superfluid does not break the translational symmetry ($\Delta^\mu_{\rm af} = 0$), but shows XY-ordering: $|\langle \hat b_1^\dagger \hat b_2 \rangle| > |\langle \hat b_1^\dagger \rangle \langle \hat b_2 \rangle |$.

\begin{figure}
\includegraphics[scale=.3]{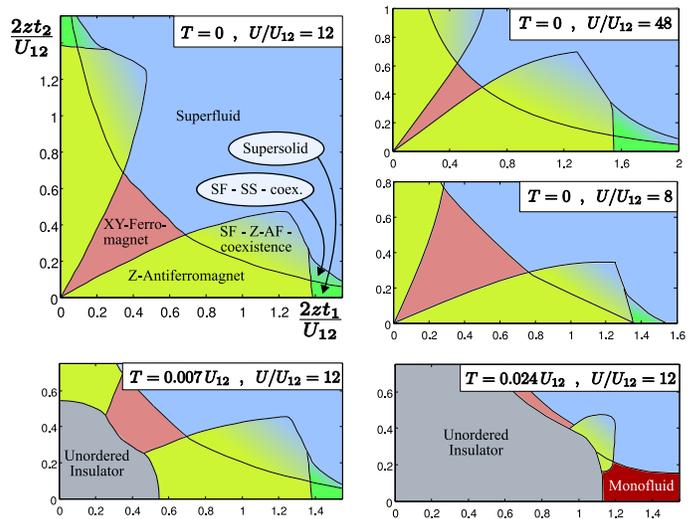}
\caption{(Color online) Phase diagram of a two-component bosonic mixture (in an optical
lattice with $z=4$) at total density $n_1 + n_2=1$
for different temperatures and ratios of the interspecies to the intraspecies interaction.}
\label{PD2comp}
\end{figure}

\begin{figure*}
\includegraphics[scale=.25]{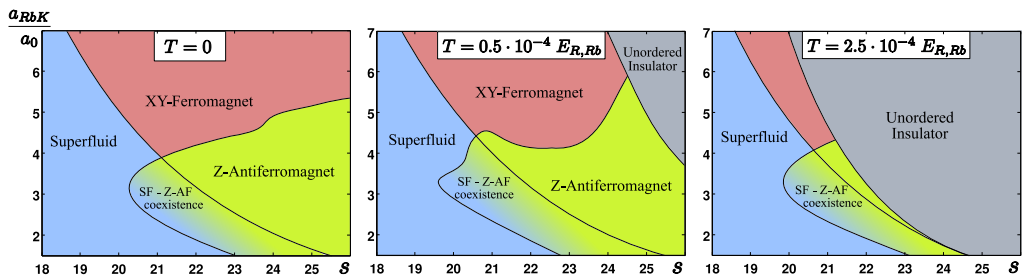}
\caption{(Color online) Phase diagram of a $^{87}$Rb-$^{41}$K-mixture (for
$z=6$) at fixed total density $n_{\rm Rb} + n_{\rm K}=1$ as a function of lattice depth $s$ and Rb-K-scattering length in units of the Bohr
radius.}
\label{FigRbK}
\end{figure*}

We now investigate this system by means of BDMFT which allows us to study the full range from weak to strong coupling and effects of finite temperature.
We first study the case that the intraspecies interactions $U=U_1 = U_2$ are equal and
much larger than the interspecies interaction $U_{12} \ll U$. We moreover vary the hopping
amplitudes $t_1$ and $t_2$. The condition $n_b + n_d = 1$ is enforced by the choice of the chemical potential $\mu_1 = \mu_2 = U_{12}/2$. This has the consequence that the relative density of the two components is changing throughout the phase diagram: the species with a higher hopping constant (the light species) has a slightly higher density in the superfluid phases of the phase diagram. In the insulating phases, on the other hand, the Mott gap protects the particle number and the relative densities are to a good approximation equal. 
Results are presented in Fig.~\ref{PD2comp} for various ratios $U/U_{12}$. For easy comparison with Ref.~\onlinecite{QMC2d} here $z=4$ is chosen. In agreement
with the fluctuation calculation \cite{Altman03} we obtain a $XY$-ferromagnetic state when
the hopping amplitudes are comparable. This $XY$-ferromagnetic domain shrinks if
$U/U_{12}$ becomes larger. For a larger difference between the hopping amplitudes there is
a first order phase transition towards a $Z$-antiferromagnetic state. The
$XY$-ferromagnetic to superfluid transition is of second order, while on the other hand the
$Z$-antiferromagnet to superfluid transition is first order, which is the reason for the
large coexistence regions where both $Z$-antiferromagnetic order and superfluidity are
stable/metastable, depending on the initial conditions. Within these metastable regions, patches with ground state spin order will nucleate into the metastable background. Therefore this phase will be susceptible to the appearance of quantum emulsions, as predicted for a one-dimensional Bose-Bose mixture \cite{emulsion}. However, this phenomenon is expected to be much more prominent in one dimension than in the three-dimensional situation we consider.

For very anisotropic hopping amplitudes, we predict the appearance of a supersolid phase. In
this supersolid the species with the smaller hopping amplitude (for convenience we take this as species 2, i.e. $t_1 \gg t_2$) is insulating: $\langle \hat b_2 \rangle = 0$, whereas the
other component (i.e. species 1) is superfluid: $\langle \hat b_1 \rangle \neq 0$. 
Moreover the complete system breaks translational symmetry by
developing a spin density wave, in which both components have alternating high and low on-site densities: $\Delta^\mu_{\rm af} > 0$ , for $\mu = 1,2$.

 Since the light species breaks the $U(1)$ and translational symmetries simultaneously, this phase is a true supersolid. The breaking of the translation symmetry is mediated by the presence of the heavy species, akin to the situation in a Bose-Fermi mixture, where supersolid order has been predicted as well\cite{Buchler03, Irakli}.
The supersolid has a first order transition to the
superfluid and a second order phase transition to the $Z$-antiferromagnet. The latter one 
can be understood as the localization transition of the light component. The supersolid phase has not been predicted in earlier studies on Bose-Bose mixtures \cite{Altman03, Isacsson05}. Only very recently it was observed in a QMC-analysis for the
two-dimensional case \cite{QMC2d}.

For nonzero temperatures additional quantum phases appear (see Fig.~\ref{PD2comp}). At low temperatures the
XY-ferromagnet and Z-antiferromagnet in low-hopping-regions develop into an unordered
Mott-state with $\langle \hat b_\mu \rangle = \langle \hat b_1^\dagger \hat b_2 \rangle = \Delta^\mu_{\rm af} = 0$.  
The coexistence regions and insulator-to-superfluid transitions remain
unaffected.
For higher temperatures the XY-insulating phase is reduced to a small strip between the
growing unordered phase and the receding superfluid. The Z-antiferromagnetic and the
AF-SF-coexistence region diminish considerably. 
For certain parameters the counter-intuitive phenomenon of reentrant superfluidity takes place: the
low-temperature antiferromagnetic phases become superfluid when the temperature is increased. This is the case because the superfluid is more stable against temperature fluctuations than the insulating $Z$-antiferromagnet.
If the temperature is increased in the region of stability of the supersolid, first the translational symmetry is restored: we obtain a phase in which only the light component is superfluid ($\langle \hat b_1 \rangle \neq 0$, $\langle \hat b_2 \rangle = 0$), but which
has no broken translational symmetry ($\Delta^\mu_{\rm af} = 0$) in contrast to the supersolid.
We call this phase \emph{monofluid}. Upon further increasing the temperature, also the remaining superfluid order is lost.

\subsection{Rubidium-Potassium mixture}
Up to now, theoretical calculations were mainly performed for the symmetric parameter choice $U_b = U_d$. However, the experimentally at present most relevant Bose-Bose mixture consisting of $^{87}$Rb and $^{41}$K generally does not have this property \cite{Minardi08, Minardi08b}. Here we choose the wavelength of the optical lattice equal to $\lambda = 757$nm, which yields equal dimensionless lattice depths $s=V_0/E_R$ for the two species. $E_R$ is the recoil energy and $V_0$ is the strength of the optical potential, which is proportional to the product of laser intensity and atomic polarizability. 
The ratio of the intraspecies interaction parameters is then fixed according to $U_{\rm Rb}/U_{\rm K} = m_{\rm K} a_{\rm Rb} / m_{\rm Rb} a_{\rm k} \approx 0.72$. The ratio of the hopping coefficients is also fixed: $t_{\rm Rb}/t_{\rm K} = m_{\rm K} / m_{\rm Rb}  \approx 0.47$. This choice of the wavelength turns out to be sufficiently anisotropic to show both $XY$-order and antiferromagnetic order, which is not possible for mixtures of different hyperfine states of the same atom. Choosing the wavelength far red-detuned like in Ref. \onlinecite{Minardi08} on the other hand, excludes the $XY$-phase from the phase diagram, but makes it possible to study the antiferromagnet and supersolid.

Experimentally, the ratio of intra-species interaction to inter-species interaction can be tuned via Feshbach-resonances \cite{Minardi08b}. Furthermore the optical lattice depth $s$ can be changed to tune the ratio $U_{\rm Rb}/t_{\rm Rb}$. We investigate the resulting $s$-$a_{\rm RbK}$ phase diagram at fixed total density $n_{\rm Rb} + n_{\rm K}=1$ by means of BDMFT, still using the semicircular DOS, but taking now lattice coordination
number $z=6$ as corresponding to the three-dimensional situation.
Results are shown in Fig.~\ref{FigRbK}.
This mixture displays 
superfluid, XY-ferromagnetic and Z-antiferromagnetic phases and we also observe the hysteresis (metastable)
region between superfluid and antiferromagnet. 
For nonzero temperatures and high lattice
depths the unordered Mott-state appears. At the temperature $T = 2.5\times10^{-4}\;E_{R,Rb} $,
($E_{R,Rb}$ being the recoil energy of Rubidium), the ordered insulating states are
reduced to a small part of parameter space, which is diminished even further for higher
temperatures. 
In order to compare this temperature to the temperatures reached in recent experiments, we consider a simple model of free bosons that undergo adiabatic time evolution while the optical lattice is ramped up. We obtain an estimate for the temperature at the relevant lattice depth in the Florence group that is one order of magnetitude larger than the highest temperature investigate here \cite{Minardi08}. Recent direct measurements of the temperature of a spinful bosonic mixture in an optical lattice at MIT have yielded temperatures which correspond to only twice the highest temperature we consider \cite{Weld09}. 

The phase diagrams in Fig.~\ref{FigRbK} are valid for the case of a shallow harmonic trap, where in the trap center the potential is very flat. Moreover, the two species need to be equally distributed in the trap center, which means that the gravitational sag has to be compensated.

\section{Conclusions} 
We have derived bosonic DMFT within a $1/z$-expansion and extended the  
formalism to
the full multi-component Bose-Hubbard model in finite dimensions. We  
first validated
the method by applying it to spinless bosons. Qualitative and  
quantitative agreement
with other methods was established. Subsequently we investigated a two- 
component mixture.
we applied the method to a two-component mixture. 
A rich phase diagram including spin-ordered
and supersolid phases was found. We furthermore calculated the experimentally relevant phase diagrams for a $^{87}$Rb- $^{41}$K in an optical lattice at zero and finite temperature.


\emph{Note added.} Recently, the BDMFT equations were also solved for the single-component Bose-Hubbard model in infinite dimensions, using a similar Anderson Hamiltonian and the same Exact Diagonalization approach as originally proposed by us \cite{Hu}.

\section*{Acknowledgements}
We thank K. Byczuk, D. Vollhardt, and F. Minardi for useful discussions. This work was supported by the German Science Foundation DFG via Forschergruppe FOR 801 and Sonderforschungsbereich SFB/TRR 49 and by the Nederlandse Organisatie voor Wetenschappelijk Onderzoek (NWO)


\begin{thebibliography}{99}
\bibitem{Fisher}
M.P.A.~Fisher, P.B.~Weichman, G.~Grinstein and D.S.~Fisher, Phys. Rev. B {\bf 40}, 546
(1989). 

\bibitem{Jaksch} 
 D. Jaksch, C. Bruder, J. I. Cirac, C. W. Gardiner, and P. Zoller, Phys. Rev. Lett. {\bf 81}, 3108 (1998).


\bibitem{Minardi08} 
J. Catani, \emph{et al.},
Phys. Rev. A {\bf 77}, 011603(R) (2008).

\bibitem{Minardi08b} 
G. Thalhammer \emph{et al.},
Phys. Rev. Lett. {\bf 100}, 210402 (2008).

\bibitem{Lukin08} 
S. Trotsky \emph{et al.},
Science {\bf 319}, 295 (2008).


\bibitem{Gutzwiller}
 D.S. Rokhsar and B.G. Kotliar, Phys. Rev. B {\bf 44}, 10328 (1991); 
 K. Sheshadri \emph{et al.}, Europhys. Lett. {\bf 22}, 257 (1993). 


\bibitem{Kuklov03} A.B. Kuklov and B.V. Svistunov Phys. Rev. Lett. {\bf 90}, 100401
(2003). 

\bibitem{Duan03}
 L.-M.   Duan, E. Demler, and M.D. Lukin, Phys. Rev. Lett. {\bf 91}, 090402 (2003).

\bibitem{Altman03} 
E. Altman {\it et al.},
New Journal of Physics {\bf 5}, 113 (2003).

\bibitem{Isacsson05} 
A. Isacsson {\it et al.},
Phys. Rev. B {\bf 72}, 184507 (2005).

\bibitem{Powell09}
S. Powell, Phys. Rev. A {\bf 79}, 053614 (2009).

\bibitem{Alon06}
O.E. Alon, A.I. Streltsov, and L.S. Cederbaum, Phys. Rev. Lett. {\bf 97}, 230403 (2006).

\bibitem{Kleine08}
A. Kleine, C. Kollath, I. McCulloch, T. Giamarchi, U. Schollwoeck, Phys. Rev. A {\bf 77}, 013607 (2008).


\bibitem{QMC2d} 
S.G. S\"oyler, B. Capogrosso-Sansone, N.V. Prokof'ev, and B.V. Svistunov,
New J. Phys. {\bf 11}, 073036 (2009). 

 \bibitem{DMFT}
 W. Metzner and D. Vollhardt, Phys. Rev. Lett. {\bf 62}, 324 (1989);  
 A. Georges {\it et al}., Rev. Mod. Phys. {\bf 68}, 13 (1996).


\bibitem{Byczuk08}
K. Byczuk and D. Vollhardt, Phys. Rev. B {\bf 77}, 235106 (2008).


\bibitem{QMC3d} B. Capogrosso-Sansone, N. V. Prokof'ev, B. V. Svistunov,  Phys. Rev. B
{\bf 75}, 134302 (2007).

\bibitem{Pelster} F.E.A.~dos~Santos and A.~Pelster, Phys. Rev. A {\bf 79}, 013614 (2009). 

\bibitem{Pelster2} M. Ohliger and A. Pelster, e-print:arXiv 0810.4399 

\bibitem{Semerjian09} G. Semerjian, M. Tarzia, and F. Zamponi, Phys. Rev. B {\bf 80}, 014524 (2009).

\bibitem{emulsion} T. Roscilde and J.I. Cirac, Phys. Rev. Lett. {\bf 98}, 190402 (2007); P. Buonsante {\it et al.}, Phys. Rev. Lett. {\bf 100},  240402 (2008). 


\bibitem{Buchler03} H. P. B\"uchler and G. Blatter, Phys. Rev. Lett. {\bf 91}, 130404 (2003).


\bibitem{Irakli} I. Titvinidze, M. Snoek, and W. Hofstetter, 
Phys. Rev. Lett. {\bf 100}, 100401 (2008).

\bibitem{Weld09} D. M. Weld, P. Medley, H. Miyake, D. Hucul, D. E. Pritchard,
and W. Ketterle, Phys. Rev. Lett. {\bf 103}, 245301 (2009).


\bibitem{Hu} W.-J. Hu and N.-H. Tong, Phys. Rev. B {\bf 80}, 245110 (2009).

\end{thebibliography}
\end{document}